\documentclass[useAMS,usenatbib]{mn2e}

\usepackage{rotating}

\newcommand{\gtappeq}{\raisebox{-0.6ex}{$\,\stackrel
{\raisebox{-.2ex}{$\textstyle >$}}{\sim}\,$}}

\title[BHk stars Globular Clusters]{Blue Hook Stars in Globular Clusters}
\author[A.Dieball et al.]{A. Dieball$^{1}$\thanks{E-mail: andrea@astro.soton.ac.uk},
  C. Knigge$^{1}$, T. J. Maccarone$^{1}$, K. S. Long$^{2}$,
  D. C. Hannikainen$^{3}$, 
\newauthor
D. Zurek$^{4}$ and M. Shara$^{4}$\\
$^{1}$School of Physics \& Astronomy, University of Southampton,
  Southampton SO 17 1BJ, UK\\
$^{2}$Space Telescope Science Institute, Baltimore, MD 21218\\
$^{3}$Mets\"ahovi Radio Observatory, Mets\"ahovintie 114, Kylm\"al\"a
  02540, Finland\\
$^{4}$Department of Astrophysics, American Museum of Natural History,
New York, NY 10024\\
}
\begin{document}

\date{Accepted 2008. Received 2008; in original form 2008}

\pagerange{\pageref{firstpage}--\pageref{lastpage}} \pubyear{2008}

\maketitle

\label{firstpage}

\begin{abstract}
Blue hook (BHk) stars are a rare class of horizontal branch stars that so
far have been found in only very few Galactic globular clusters (GCs). The
dominant mechanism for producing these objects is currently still
unclear. In order to test if the presence of BHk populations in
a given GC is linked to specific physical or structural
cluster properties, we have constructed a parent sample of GCs for
which existing data is sufficient to establish the presence or absence
of BHk populations with confidence. We then compare the
properties of those clusters in our parent sample that do contain a
BHk population to those that do not. We find that there is only
one compelling difference between BHk and non-BHk
clusters: all known BHk clusters are unusually massive. However,
we also find that the BHk clusters are consistent with being
uniformly distributed within the {\it cumulative} mass distribution of
the parent sample. Thus, while it is attractive to suggest there is is
a lower mass cut-off for clusters capable of forming BHk stars,
the data do not require this. Instead, the apparent preference for
massive clusters could still be a purely statistical effect:
intrinsically rare objects can only be found by searching a
sufficiently large number of stars. 
\end{abstract}


\section{Introduction}
The horizontal branch (HB) in globular clusters (GCs) represents the
Helium core burning phase of stellar evolution. HB morphology varies
from cluster to cluster; some clusters show a red HB (RHB), others a
blue HB (BHB), and some even a bimodal HB with both a RHB and BHB. 
Some BHBs exhibit a long, vertically extended tail in optical
colour-magnitude diagrams (CMDs), which is formed by particularly hot
objects. BHB stars hotter than $T_{eff} \approx 20000$ K are usually
referred to as extreme HB (EHB) stars and are thought to 
have undergone severe mass loss during their RGB phase (e.g. Momany et
al.\ 2004). They evolve into AGB manqu{\'e} stars or post-early AGB
stars, but do not return to the AGB (e.g. Dorman et al.\ 1993). In
a few GCs, the hottest -- and optically faintest -- end of the
EHB at $T_{eff} > 31500$ K is populated by a peculiar class of objects
which was first detected in $\omega$ Cen and NGC\,2808. They form a
blue hook (BHk) at the hot end of the HB in far-UV (FUV) CMDs and were
consequently called ``blue hook stars'' (Whitney et al.\ 1998, D'Cruz
et al.\ 2000, Brown et al.\ 2001).   

The physical mechanism that produces BHk populations is still
uncertain. At least two scenarios have been proposed.   

In the {\it late He flasher scenario} these stars are explained as a
consequence of extreme mass-loss during the RGB phase and late
He-flashing while descending the white dwarf (WD) cooling sequence 
(e.g.\ Brown et al.\ 2001). Due to the thin residual H-envelope, He is
mixed into the envelope and H is mixed into the core during the late
He-flash. As a result, the stars are hotter and UV-fainter than
canonical EHB stars. 

By contrast, in the {\it He self-enrichment scenario} the EHB
and BHk stars are produced via the normal evolution of He-enriched
sub-populations in GCs. For example, Lee et al.\ (2005) were able to
explain the peculiar HB morphology and the presence of BHk stars in
NGC\,2808 and $\omega$ Cen with several He-enhanced sub-populations
within these clusters. These sub-populations might have formed from
the ejecta of intermediate-mass AGB stars of the first generation of
stars (see also D'Antona et al.\ 2005). For the same age and
metallicity, He-enriched HB stars have smaller masses than normal HB
stars, resulting in bluer ZAHB locations (Lee et al.\ 1994). They are
also brighter in the FUV, but this effect is reversed for
very hot He-enriched HB stars with $T_{eff} > 19000$ K. Lee et
al.\ (2005) also predicted a narrow split in NGC\,2808's main sequence
(MS), which was indeed found by D'Antona et al.\ (2005) and Piotto et
al.\ (2007).    

One way to make progress in understanding the nature of BHk stars and
their progenitors is to test if the presence of BHk populations is
associated with particular cluster properties. For example, if
He-enriched populations are needed to produce BHk stars, 
these stars would then only be expected in GCs that exceed some
critical mass threshold needed to keep the stellar ejecta from the
first generation of stars within the cluster potential. Alternatively,
if He enrichment is not important in the production of BHk stars, the
mass loss required by the late-He flash could be due to dynamical
processes, such as binary interactions. In this case, one might expect
that only clusters with high encounter rates host BHk populations.   

In order to address these questions, we have
constructed a sample of GCs in which the presence or absence of BHk
populations can be established with a high degree of confidence. More
specifically, since BHk stars can (only) be easily recognized in UV
CMDs, we restrict our sample to GCs for which sufficiently deep
(spacebased) FUV or near-UV (NUV) observations
are available. Using this sample, we are able to
search for any correlations between the existence of BHk populations
and the properties of their host clusters (e.g. metallicity, mass,
relaxation times etc.). Our results do confirm suggestions in the 
literature that BHk stars are preferentially found in massive clusters
(e.g.\ Rosenberg et al.\ 2004, Rood et al.\ 2008). However, we 
also show that this preference may actually just be a selection
effect, associated with the fact that BHk stars are intrinsically rare
and thus can only be found in sufficiently large samples of stars. 

\section{The cluster sample}

\label{data}
\begin{figure}
\includegraphics[width=8.7cm,height=5.7cm]{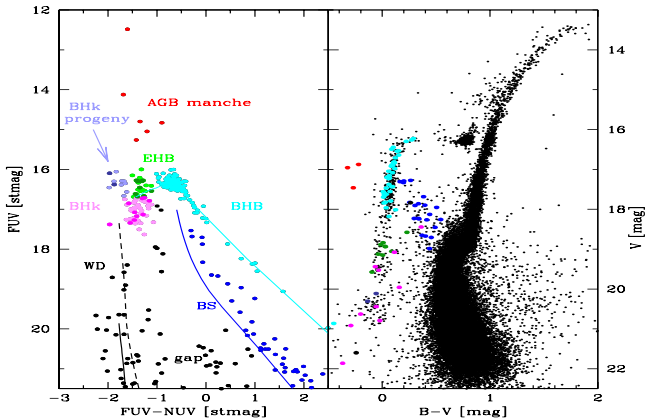}
\caption{Left: FUV CMD of NGC\,2808. BSs are marked in blue, WD
  candidates and gap sources/CV candidates in
  black, BHB stars in cyan, EHB stars in green ($-1.5 < FUV - NUV <
  -1.1$, $16.7 < FUV \ge 16.0$), AGB manqu{\'e} stars in red ($12.5 <
  FUV < 15.3$ ), BHk candidates in magenta ($FUV - NUV < -1$, $17.65 <
  FUV < 16.7$), and BHk progeny stars in violet ($FUV - NUV < -1.6$,
  $16.1 < FUV < 16.7$). EHB and BHk stars and their progeny 
  that have an optical counterpart are marked in a darker
  shade. Right: optical CMD of NGC\,2808 (data taken from Piotto et
  al.\ 2002). Counterparts to FUV sources are marked with the same
  colour as in the left CMD. \label{cmd2808}}
\end{figure}

\setlength{\tabcolsep}{0.05cm}
\begin{table*}
\begin{center}
\caption{
  Properties of the Globular Cluster sample that were searched for BHk
  stars. The distance, reddening $E_{B-V}$, $[\rm{Fe/H}]$, specific RR
  Lyrae frequency S(RR), HB ratio HBR, the core, half-mass and tidal
  radii (columns 2 -- 8), and the ellipticity, the logarithmic core
  and half-mass relaxation times, and the logarithmic central luminosity density
  (columns 11 -- 14) are taken from Harris (1996). The concentration
  parameter $c = lg(r_{tidal}/r_{core})$ is given in col 10. The
  cluster mass and the central and half-mass escape velocities 
  (cols 15 -- 17) are taken from Gnedin et al.\ (2002). The collision
  rate $\Gamma =  \propto \rho_{c}^{1.5} \cdot r_{c}^2$ (Verbunt \&
  Hut 1987, Heinke et al.\ 2003, Fregeau 2008), scaled with
  $\Gamma_{47Tuc}$, is given in column 18; columns 19 -- 22 indicate
  whether the cluster contains a BHB, EHB, BHk or multiple population,
  column 23 gives a lower limit to the number of BHk stars estimated
  from available CMDs. References for the BHk stars are 
  (a) Brown et al.\ (2001), (b) Busso et al.\ (2004, 2007; the
  NGC\,6441 CMD shows a large scatter, we overplotted our own ZAHB
  models and found maybe four (less certain) candidates), (c)
  Connelly et al.\ (2006), (d) D'Alessandro et al. (2008), (e) D'Cruz
  et al.\ (2000), (f) Dieball et al.\ (2007), (g) Dieball et al.\ (in
  prep.), (h) Ferraro et al.\ (1998), (i) Hill et al.\ (1996), (j)
  Knigge et al.\ (2002), (k) Landsman et al.\ (1996),
  (l) M\"ohler et al.\ (1997, one spectroscopically confirmed BHk
  star, other than that no BHk population is visible), (m) M\"ohler et
  al.\ (2004, 2007), (n) Mould et al.\ (1996), (o) Parise et
  al.\ (1998), (p) Ripepi et al.\ (2007), (q) Rosenberg et 
  al.\ (2004), (r) Sandquist \& Hess (2008), (s) Watson et al.\ (1994,
  one subluminous HB star reported, possibly a BHk candidate, but
  other than that no BHk population was detected), (t) Zurek et
  al.\ (in prep.). (D) in column 22 indicates that the cluster
  was discussed in D'Antona \& Caloi (2008) who suggest that the cluster
  contains a sizeable population of a second generation of stars.}   
\label{table}   
\vspace{0.2cm}
\tiny
\begin{tabular}{lcccccccccccccccccccccc}
1 & 2 & 3 & 4 & 5 & 6 & 7 & 8 & 9 & 10 & 11 & 12 & 13 & 14 & 15 & 16 &
17 & 18 & 19 & 20 & 21 & 22 & 23\\
cluster&dist&$E_{B-V}$&$[\rm{Fe/H}]$&S(RR)&HBR&$r_{c}$&$r_{hm}$&$r_{tidal}$&$c$&$e$&$lg(t_{c})$&$lg(t_{hm})$&$lg(\rho_{c})$&$M_{tot}$&$v_{c}$&$v_{hm}$&$\Gamma$& BHB & EHB & BHk & mp & $\#$BHk \\
 &[kpc]&[mag]& & & &[pc]&[pc]&[pc]& & & & & &[$10^{6} M_{\odot}$]&[km/s]&[km/s]& & & & & \\
\hline
NGC2419    &84.2&0.11&-2.12& 4.6&0.86 &8.57&17.88&214.07&1.40&0.03&9.96&10.55&1.54&1.74&27.8 &19.6&0.003&y&y?&y$^{(d,p,r)}$&  &$\gtappeq 190$\\
NGC2808    &9.6 &0.22&-1.15& 0.3&-0.49&0.73&2.12 &43.42 &1.77&0.12&8.30& 9.13&4.59&1.42&72.8 &45.8&0.912&y&y &y$^{(a,m)}$  &y &$\gtappeq 50$ \\    
$\omega$Cen& 5.3&0.12&-1.62&12.3&     &2.16&6.44 &87.93 &1.61&0.17&9.05&10.00&3.37&3.35&60.4 &44.0&0.119&y&y &y$^{(e,m)}$  &y &$\gtappeq 25$ \\
NGC6388    &10.0&0.37&-0.61& 2.4&     &0.35&1.95 &18.06 &1.70&0.01&7.74& 9.08&5.34&2.17&124.0&80.2&2.810&y&y &y$^{(b)}$    &HB&$\gtappeq 10$ \\ 
M54        &26.8&0.15&-1.59& 6.1& 0.75&0.86&3.82 &58.24 &1.84&0.06&8.46& 9.62&4.58&2.59&84.5 &51.6&1.230&y&y &y$^{(q)}$    &y &$\gtappeq 30$ \\ 
\hline                                                                                                                  
47Tuc      &4.5 &0.04&-0.76& 0.2&-0.99&0.52&3.65 &56.11 &2.03&0.09&7.96& 9.48&4.81&1.50&68.8 &38.0&1.000&n&n &n$^{(j)}$    &(D)&\\ 
NGC1851    &12.1&0.02&-1.23&13.5&-0.36&0.21&1.83 &41.18 &2.32&0.05&6.98& 8.85&5.32&0.55&51.8 &24.3&0.958&y&n &n$^{(t)}$    &y  &\\   
M79        &12.9&0.01&-1.58& 2.2& 0.89&0.60&3.00 &31.30 &1.72&0.01&7.78& 9.10&4.00&0.36&40.7 &26.1&0.081&y&y &n$^{(i)}$    &   &\\
M3         &10.4&0.01&-1.57&49.0& 0.08&1.66&3.39 &115.54&1.84&0.04&8.84& 9.35&3.51&0.96&37.2 &22.7&0.115&y&n &n$^{(h)}$    &(D)&\\
M80        &10.0&0.18&-1.77& 3.1& 0.93&0.44&1.89 &38.63 &1.95&0.00&7.73& 8.86&4.76&0.50&48.7 &28.1&0.592&y&y &n$^{(g,h)}$  &   & 1?\\  
M13        &7.7 &0.02&-1.54& 1.7& 0.97&1.75&3.34 &56.40 &1.51&0.11&8.80& 9.30&3.33&0.78&39.1 &26.9&0.068&y&y &n$^{(h,o)}$  &(D)&\\
NGC6441    &11.7&0.47&-0.53& 0.3&     &0.37&2.18 &27.23 &1.85&0.02&7.77& 9.19&5.25&1.57&102.0&62.0&2.364&y&y &n?$^{(b)}$   &HB & 4?\\ 
M70        &9.0 &0.07&-1.52& 2.9& 0.96&0.08&2.44 &20.71 &2.50&0.01&5.62& 8.83&5.41&0.18&39.3 &17.3&0.183&y&y &n?$^{(c,s)}$ &   & 1?\\
NGC6752    &4.0 &0.04&-1.57& 0.0& 1.00&0.20&2.72 &64.40 &2.50&0.04&6.83& 9.01&4.91&0.32&32.9 &14.5&0.205&y&y &n$^{(k)}$    &   &\\
M15        &10.3&0.10&-2.27&18.9& 0.67&0.21&3.18 &64.42 &2.50&0.05&7.02& 9.35&5.38&1.19&62.1 &27.4&1.168&y&n &n?$^{(f,l)}$ &(D)& 1\\ 
NGC7099    &8.0 &0.03&-2.12& 3.2& 0.89&0.14&2.68 &42.68 &2.50&0.01&6.38& 8.95&5.04&0.24&34.1 &15.0&0.160&y&n?&n$^{(n)}$    &   &\\
\hline
\end{tabular}
\end{center}
\end{table*}
\normalsize

As the nomenclature of BHB, EHB and BHk stars is often unclear, and 
various authors might in fact talk about
different stellar populations, it is essential to first make a clear
definition of what we call EHB and BHk stars. This is best illustrated
in the FUV CMD of NGC\,2808 (Fig.~\ref{cmd2808} left panel, taken from
Dieball et al.\ 2005\footnote{For orientation purposes, theoretical
  tracks are included in this plot. The ZAHB was constructed based on the
  BaSTI ZAHB models (e.g. Cordier et al.\ 2007). Note that in Dieball
  et al.\ (2005), we used the Dorman (1992) model for the ZAHB. For
  further details on the CMD and the theoretical tracks, see Dieball
  et al.\ (2005).}). Various populations show up in the CMD and are
marked with different colours, their corresponding counterparts are
marked with the same colour in the optical CMD (Fig.~\ref{cmd2808}
right panel, taken from Piotto et al.\ 2002). 
In this paper, we call those FUV sources {\bf EHB stars} that are at least
as blue as the ZAHB at $T_{eff} > 20000$ K (following
e.g. Momany et al.\ 2004). Stars that are fainter than
the EHB stars (i.e. below the ZAHB) but which have a similar colour
in the FUV CMDs are the {\bf blue hook candidates} (see also Brown et
al.\ 2001). Stars brighter in FUV than the EHB stars might be 
{\bf AGB   manqu{\'e} stars}. We see six such stars in our 
Fig.~\ref{cmd2808}. The small population of stars bluer than the EHB
stars and brighter than the BHk stars agree with being {\bf blue hook
  progeny}.    

BHk stars were first detected in $\omega$ Cen and NGC\,2808 through
the subluminous blue hook that these stars form at the hot end of the
HB in the FUV CMDs. As Brown et al. (2001) point out, the higher
$T_{eff}$ of BHk stars compared to the cooler EHB
stars imply larger bolometric corrections in the optical and hence
fainter optical magnitudes. As a result, the identification of BHk
stars in optical CMDs is much more difficult. In constructing a parent
sample of GCs for statistical purposes, the key requirement is that
BHk populations, if present, could be identified with confidence. As
already noted above, we therefore mainly restrict our sample to
GCs for which sufficiently deep (spacebased) UV data exist. We allow
only two exceptions to this rule: M\,54 and NGC\,2419. 
BHk stars have been suggested in both of these clusters based on deep
optical data, and we find these studies convincing. Our final parent
sample is listed in Table~1, along with the relevant physical
parameters for each cluster. 
  
\section{Analysis}

\label{discussion}

\begin{figure}
\includegraphics[width=8.7cm,height=6cm]{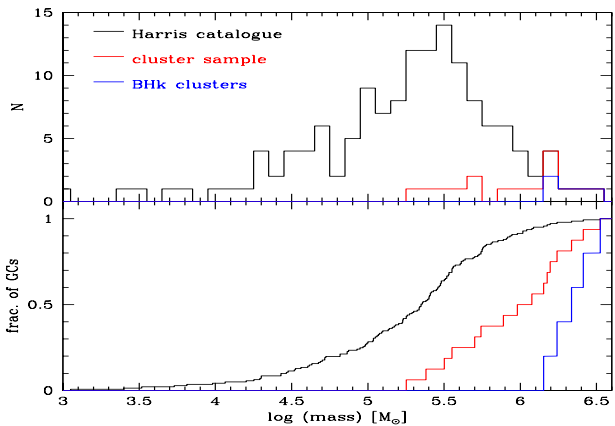}
\caption{Top panel: Histogram of the logarithmic mass distribution of
  all clusters in our sample (red line), the BHk clusters (blue
  line), and all clusters in the Harris catalogue (black line). Bottom
  panel: Cumulative mass distribution of the sample clusters (red),
  the BHk clusters (blue), and the Harris catalogue (black). \label{mass}}  
\end{figure}

We searched for statistically significant differences between the 
parameters of the five BHk clusters and the 11 non-BHk clusters in the 
parent sample using a series of Kolmogorov-Smirnov (KS) tests. The KS
test returns the probability that the difference between two populations
should be as large as observed under the null hypothesis that both
distributions are drawn from the same parent distribution. The results
of the KS tests are given in Table~\ref{KS}. For reference, we have
also carried out KS tests comparing our parent sample to the Harris
(1996) catalogue of GCs in the Milky Way.

\begin{table}
\caption{Probabilities returned from the KS test, given in
  \%. We compare the various parameters (see Table~\ref{table}) of the
  non-BHk clusters in our sample with the BHk clusters (column 2), and
  our parent sample including the BHk clusters with the Harris (1996)
  catalogue (column 3). \label{KS}}     
\begin{tabular}{l|c|c}
               & non-BHk vs. BHk & parent vs. Harris \\           
\hline
$M_{tot}$       & 0.79  &  0.04  \\  
$c$            & 2.52  &  0.46  \\   
$e$            &46.48  & 27.56  \\   
$[\rm{Fe/H}]$  &76.77  & 26.23  \\   
S(RR)          &76.77  & 26.20  \\   
HBR            &24.23  & 16.10  \\
$r_{c}$         & 8.48  &  8.77  \\   
$r_{hm}$        &10.19  & 42.58  \\   
$r_{tidal}$     &70.72  &  4.89  \\   
$lg(t_{c})$     & 8.48  & 40.22  \\   
$lg(t_{hm})$    &10.19  & 18.59  \\   
$lg(\rho_{c})$  &20.05  &  1.12  \\   
$v_{c}$         &20.05  &  0.04  \\   
$v_{hm}$        & 3.13  &  0.50  \\   
$\Gamma$       &82.45  &  0.09  \\
\hline
\end{tabular}
\end{table}

Table~2 shows that only very few statistically significant
correlations exist between the occurrence of BHk stars and the
parameters of the clusters. As can be seen, the most significant
difference between BHk and non-BHk clusters is associated with their
mass distributions. More specifically, the KS test indicates at
greater than 99\% confidence that (high) mass is associated with the
presence of BHk stars in a cluster. In Fig.~\ref{mass} we plot the
mass distribution and the cumulative mass distribution for the BHk
clusters (blue line), all clusters in our sample (red line), and all
clusters from the Harris (1996) catalogue. As can be seen, all
clusters in which BHk candidates have been found are amongst the most
massive in the Galaxy. The apparent preference for BHk populations to
be found in abnormally massive clusters has been remarked upon before
in the literature (see e.g. Rosenberg et al.\ 2004, M\"ohler et
al.\ 2004, Rood et al.\ 2008). At first sight, this result would seem
to support the self-enrichment scenario for BHk star production. 

However, although BHk stars appear to be found more frequently in massive 
star clusters, this does not reveal much about their origin. In
particular, there is an obvious selection bias resulting from the
fact that massive clusters contain more stars. Since studies of
massive clusters thus typically also inspect more stars, such studies
are more likely to turn up intrinsically rare populations, such as BHk
stars. If, for example, only 1 in every 100,000 stars becomes
a BHk object, individual low-mass clusters will not contain any BHk
populations, even if there is no physical lower mass limit associated
with BHk production.  

The way to test for the existence of a critical mass limit in this
type of analysis is to consider the location of BHk clusters in the sorted, 
{\em cumulative} mass distribution of the parent sample. If there is
no critical mass limit, BHk clusters should be uniformly distributed
within this cumulative mass distribution. If there is a mass limit,
they should be significantly concentrated at the high mass end. 
By doing so, we are effectively combining many low-mass clusters to
form aggregate high-mass clusters.
\footnote{The same test was used by Verbunt \& Hut (1987) to establish the
connection between bright low-mass X-ray binaries (LMXBs) and cluster
collision rates. More specifically, they showed that LMXB-hosting
clusters were uniformly distributed within the cumulative collision
rate distribution of the Galactic GC sample.}

\begin{figure}
\includegraphics[width=8.7cm,height=1.4cm]{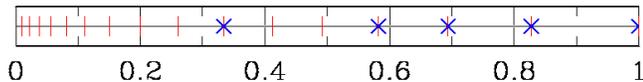}
\caption{Positions of the BHk clusters (blue) within the (normalized)
  cumulative mass distribution of all clusters (red) in our sample.
  If BHk stars scale with cluster mass, the BHk clusters should be
  evenly distributed. \label{uniform}}
\end{figure}

In Fig.~\ref{uniform} we plot the positions of the BHk clusters
within the cumulative mass distribution of the available parent
sample. Using a KS test, we then checked whether the positions of the
BHk clusters is different from a uniform distribution. The returned
probability that the BHk clusters are randomly selected from the
parent population is 36.9\%. This is not statistically significant at
even the $1\sigma$ level of confidence. 
Thus, the available sample does {\it not} provide evidence for a lower
mass limit for clusters capable of producing BHk stars.  

\section{Discussion}

Our main result is that BHk populations have so far been detected only
in the most massive GCs, but this does not (yet) imply that low-mass
clusters cannot form BHk stars. In particular, we cannot rule out that
the apparent preference for massive GCs arises simply because
these clusters contain more stars. However, regardless of whether the
sizes of BHk populations simply scale with total cluster mass or
whether a lower-mass cut-off for clusters capable of producing BHk
stars exists, cluster mass is clearly a key parameter.
That said, it is puzzling that there {\it are} GCs as massive as
the BHk clusters that do {\it not} harbour a population of BHk
stars. For example, 47\,Tuc is also one of the most massive clusters,
but does not contain any BHk stars (or BHB or EHB stars). Similarly,
NGC\,6441, an HB twin to NGC\,6388 and more massive than NGC\,2808,
also does not show a population of BHk stars. Therefore mass cannot be
the only parameter that is associated with the production of BHk stars.  

Other parameters for which the KS test suggested marginally
significant differences between the BHk and the non-BHk clusters are
the concentration parameter, the core and halfmass radius, the core
and halfmass relaxation time, and the halfmass escape velocity. It is
difficult to know whether these secondary correlations are physical,
since most of the relevant cluster parameters also correlate 
with total cluster mass within our parent sample. If one is 
nevertheless willing to take these secondary correlations at face
value, they might indicate that binarity may be a key ingredient in
BHk star formation. More specifically, the long relaxation times, large
core radii, small concentration parameters and large escape velocities
of the BHk clusters might indicate that cluster core collapse was/is
decelerated or prevented, most likely by the presence of binaries 
(e.g.\ Elson et al.\ 1987, Hut et al.\ 1992, Fregeau 2008).  
Although EHB stars seem to be rarely found in binaries
(e.g. Moni Bidin et al.\ 2008) they still might have formed in 
binary systems and could be the result of binary evolution, much like
their field counterparts, the subdwarf B (sdB) stars (e.g. Han et
al.\ 2007). Binary interaction might then enhance mass-loss
during the RGB phase of a star, finally leading to the evolution into
a late He-flasher/BHk star (e.g. Bailyn et al. 1992). On the other hand,
short relaxation times might imply that the hardening of binaries
through binary interactions (Heggie 1975) occurs so rapidly that all of
the stellar envelope is stripped while the star is still 
on the RGB, leaving behind a He WD rather than an EHB or BHk star.

Interestingly, most BHk clusters show multiple stellar populations
($\omega$ Cen, NGC\,2808, M\,54) and/or unusual and tilted HBs
(NGC\,6388) that indicate He self-enrichment (see e.g.\ Lee et al.\ 1999,
Bedin et al.\ 2004, D'Antona et al.\ 2005, Piotto et al.\ 2005,
Siegel et al. 2007, Piotto et al.\ 2007). This may be linked to the
high masses of the BHk clusters, as only clusters massive enough
should be able to retain the ejecta of their AGB stars. However, so
far no evidence for multiple, He-enriched stellar populations has been
found in NGC\,2419 (Sandquist \& Hess 2008).       

The primary result of our analysis is that from the available data one
cannot conclude that a low mass cutoff to BHk populations
exists. However, absence of evidence does not constitute evidence of
absence. Such a low-mass cut-off
may exist, and given that 4 of the 5 BHk clusters are the most massive
in our sample, the BHk sub-sample could hardly be more biased towards
high mass than it is. Thus the real problem is that the available sample
is too small, and too deficient in lower mass clusters, to permit a
more definitive test of the critical mass hypothesis. 
The available sample is itself strongly biased towards massive
clusters (see the comparison of the parent sample to the total
Galactic population in Table~2), but a larger more unbiased sample
does not exist today. This is because massive clusters usually make
more promising observational targets, for a variety of technical and
physical reasons. However, this type of selection bias severely limits
our ability to carry out sufficiently powerful statistical tests. 

\subsection{Summary}

BHk stars seem to be a rare phenomenon that occurs in only a few
GCs. We have constructed a carefully selected parent sample of GCs in
which the presence or absence of BHk populations can be established
with confidence. 
BHk stars have so far been found in only the most massive
GCs. However, we also find that this does not (yet) imply the
existence of a lower limit on the mass of a cluster capable of
producing BHk stars. Instead, the preference for massive clusters 
could just be a selection effect, associated with the fact that BHk
stars are intrinsically rare and thus it is more likely to observe
these stars in high mass clusters rather than low mass clusters. In
order to resolve this issue, we need additional data on many more 
low-mass clusters. 

Alternatively, it may be possible to make progress by studying how the
{\it numbers} of BHk stars scale with cluster properties. However,
given the non-uniformity of the existing dataset, we can only estimate
lower limits based on the published data (see Table~1).
In any case, the identification of {\it high} cluster mass as a
necessary ingredient for the production of BHk populations appears to
be premature. 

We also note that not {\it all} massive clusters, or all clusters with
multiple populations, also contain BHk stars. Counter examples are the
massive GCs 47\,Tuc and NGC\,6441. On the other hand, NGC\,2419
contains BHk stars, but apparently no multiple populations (see 
Sandquist \& Hess, 2008). Therefore we conclude that mass might be the
main, but not the only parameter relevant to establishing BHk
populations in GCs. With this in mind, we have sketched a scenario for
BHk formation involving dynamical interactions between binary systems
that would also be consistent with the fact that BHk clusters tend to
have long relaxation times, large core radii, and large escape velocities.

\section*{Acknowledgments}
DH gratefully acknowledges Academy of Finland project number 212656.

\label{lastpage}

\end{document}